\documentclass{article}

\usepackage[english]{babel}
\usepackage{graphics}
\usepackage{bm}
\usepackage{graphicx}
\usepackage{authblk}
\usepackage{epstopdf}

\title{Clusters in Separated Tubes of Tilted Dipoles}
\author{Jeremy R. Armstrong \\
  Department of Physics and Astronomy, University of Nebraska at Kearney, Kearney, NE 68849, USA\\
  \and
  
  Aksel S. Jensen \\
  Department of Physics and Astronomy, Aarhus University, DK-8000 Aarhus C, Denmark\\
  \and
  Artem G. Volosniev\\
  Institute of Science and Technology Austria, 3400 Klosterneuburg, Austria\\
  Institute for Nuclear Physics, Technical University Darmstadt, 64289 Darmstadt, Germany\\
  \and
  Nikolaj T. Zinner\\
  Aarhus Institute of Advanced Studies, Aarhus University, DK-8000 Aarhus C, Denmark\\
  Department of Physics and Astronomy, Aarhus University, DK-8000 Aarhus C, Denmark}

\begin{document}
\maketitle

\begin{abstract}
  A few-body cluster is a building block of a many-body
system in a gas phase provided the temperature at most is of the
order of the binding energy of this cluster.  Here we illustrate this statement by
considering a system of tubes filled with dipolar distinguishable
particles. We~calculate the partition function, which determines the
probability to find a few-body cluster at a given temperature. The
input for our calculations---the energies of few-body clusters---is estimated using the harmonic approximation.  We first describe
and demonstrate the validity of our numerical procedure.  Then we
discuss the results featuring melting of the zero-temperature
many-body state into a gas of free particles and few-body
clusters.  For temperature higher than its binding energy threshold,
the dimers overwhelmingly dominate the ensemble, where the remaining
probability is in free particles.  At very high temperatures free
(harmonic oscillator trap-bound) particle dominance is eventually
reached.  This structure evolution appears both for one and two
particles in each layer providing crucial
information about the behavior of ultracold dipolar gases.  The investigation addresses the transition region between few- and many-body physics as a function of temperature using a system of ten dipoles in five tubes.
\end{abstract}


\section{Introduction}

One important question that quantum few-body physics should answer is under which
conditions few-body bound states play a role (or could be observed) in a many-body system.
It is clear that if the energy associated with the temperature is much larger than the few-body binding energy, then bound states occupy a tiny fraction of the Hilbert space, and~hence the probability to observe (populate) a bound state is exponentially suppressed. Think, for~example, about Efimov states~\cite{efimov1970} (see~References~\cite{jensen2004,braaten2006,greene2017,naidon2017} for a review) in cold-atom systems. These states are always present in a cold gas; however, only at ultracold temperatures is it possible to observe them~\cite{kraemer2006}.

\begin{figure}
\centering
\includegraphics[width=0.95\columnwidth]{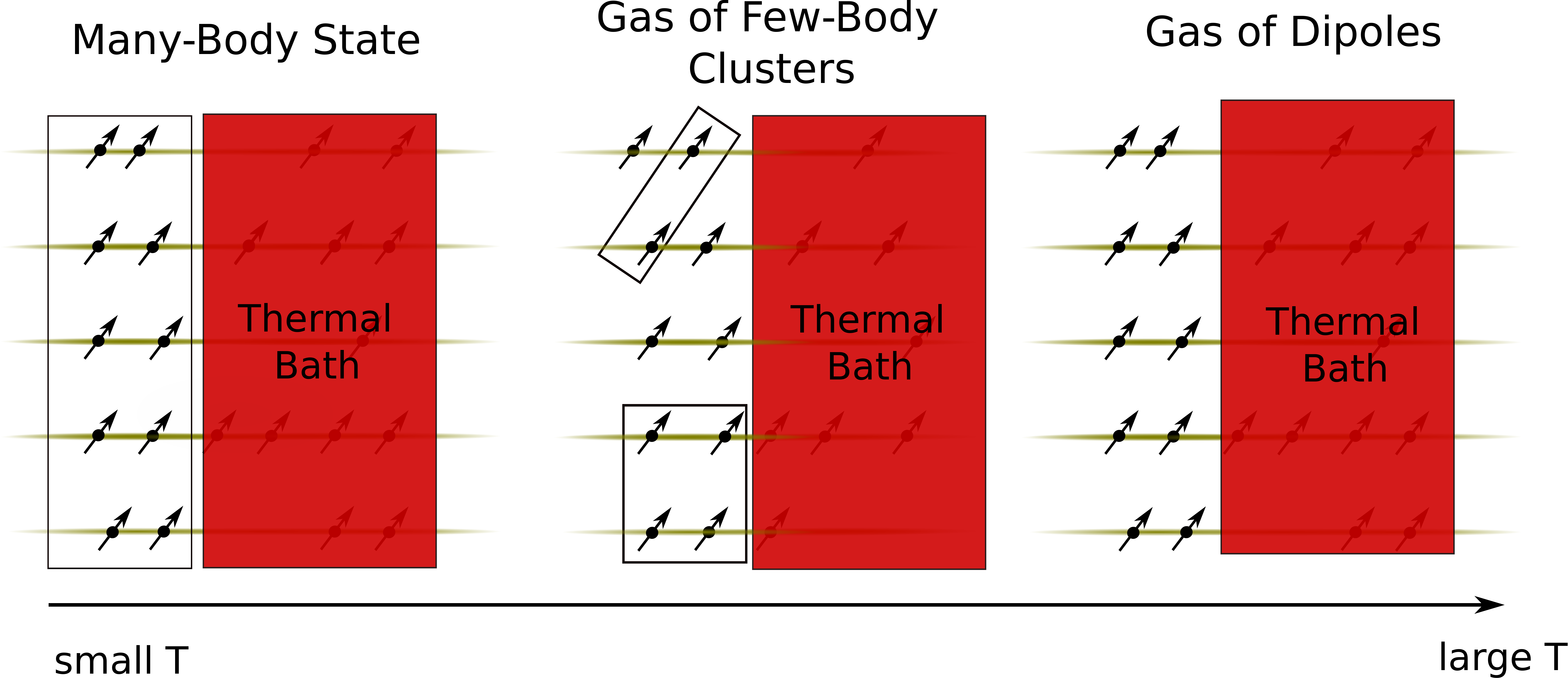}
\caption{The system of interest is five one-dimensional tubes filled with two dipolar particles with dipole moments aligned at the so-called ``magic angle''. The~system is in a thermal equilibrium with a~bath at temperature $T$. At~high temperatures, the~system will consist of a gas of independent particles, and~at zero temperature the attraction between the layers will lead to a certain bound structure.  At~intermediate temperatures, various few-body clusters will~form.  }
\label{Figure1}
\end{figure}

In this paper, we study at which temperatures few-body bound states can be observed in a~cold gas of dipoles (see~\cite{lahaye2009, baranov2012}, which review cold dipolar gases), { once precise control of cold polar molecules is achieved~\cite{bohn2017}.} Our model is the system illustrated in Figure~\ref{Figure1}. The~dipoles are trapped by an optical lattice, which can be formed, for~example, by~superimposing two orthogonal standing  waves~\cite{bloch2008}. Strong trapping prevents particles from tunneling between the tubes, so the system can be approximated as a collection of one-dimensional tubes. An~external electric field controls the alignment of the dipoles. Previous works investigated the formation of chains in two-dimensional geometries~\cite{wang2006}. We~are interested in formation of few-body states with more than one particle per layer (or tube), which are unlikely to be observed for dipoles with perpendicular polarization~\cite{volo13}. Therefore, we~assume that the dipoles are tilted to the ``magic angle''  such that there is no interaction within a~tube, see, References~\cite{wunsh2011, volo13} for a~discussion of few-body bound states with other polarizations. Still~the long-range dipole-dipole interaction allows particles to interact between the~layers. This~interaction supports a~zoo of few-body bound states~\cite{wunsh2011,zinner2011,volo13,bjerlin2018}, whose presence should be taken into account when building models of the corresponding many-body systems (see, for~example,~\cite{wang2006,pikovski2010,dalmonte2011, kuklov2011,cinti2017}).

To find at which temperature these states enter the description, we
consider a system of dipoles coupled to a thermal bath. We~assume that particles obey
Boltzmann statistics, { however, we~will argue that our results are also applicable to systems of bosons.} For the sake of discussion, we~assume
that the system is made of ten dipoles that occupy five tubes, see
Figure~\ref{Figure1}. In~spite of its simplicity, we~expect that this
system contains all basic ingredients allowing us to learn about the
formation of the simplest few-body clusters. { This system has enough tubes so that  particles in the outermost tubes do not interact with each other. Therefore, adding more tubes cannot qualitatively change our results. Moreover, the~system has more than one particle per tube allowing us to investigate the effect of non-chain few-body structures. Our results show that these structures are not important for our analysis. In~particular, our results show that there is a~clear transition from a many-body bound state to a state dominated by chains of dipoles.}

This paper is organized as follows. In~Section~\ref{subsec:energ}, we~introduce the method used for computation of few-body energies. The~partition function that determines the probability of a~particular state is discussed in Section~\ref{subsec:part}.  In~Section~\ref{sec:Res}, we~demonstrate at which temperatures few-body clusters can be observed. In~Section~\ref{sec:concl}, we~summarize our results and~conclude.


\section{Binding Energies of~Clusters}
\label{subsec:energ}

The binding energy of a specific cluster can be obtained by diagonalizing the Hamiltonian
\begin{eqnarray}
H^{dd}&=&-\frac{\hbar^2}{2m}\sum_{i,\alpha}\frac{d^2}{dx_{i,\alpha}^2}+\sum_{i,j;\alpha>\beta}V_{dip}(x_{i,\alpha}-x_{j,\beta}),
\label{eq3}
\end{eqnarray}
where $m$ is the mass of the dipolar particle, the~subscript $\{i,\alpha\}$ refers to the $i$th particle in the $\alpha$th layer.  The~potential, $V_{dip}$, describes the dipole-dipole interaction:
\begin{equation}
V_{dip}(x_{i,\alpha}-x_{j,\beta})=\frac{\mathbf{D}_{i,\alpha}\mathbf{D}_{j,\beta}r^2-3(\mathbf{D}_{i,\alpha}\mathbf{r})(\mathbf{D}_{j,\beta}\mathbf{r})}{r^{5}},
\label{eq2a}
\end{equation}
where $\mathbf{r}=(x_{i,\alpha}-x_{j,\beta},0,n d)$ is the relative distance between the two dipoles, $n=\alpha-\beta$ determines the separation between the dipoles in the $z$ direction, $d$ is the distance between the adjacent tubes and $\mathbf{D}_{i,\alpha}$ is the dipole moment of the $i$th dipole in the $\alpha$th layer. For~simplicity, the~width of a tube is taken to be zero (for finite widths, see References~\cite{sinha2007, reimann2010, zinner2011}). By~assumption, the~dipole moment has only $x$ and $z$ components: $\mathbf{D}_{i,\alpha}=D_{i,\alpha}(\cos(\phi),0,\sin(\phi))$. Our choice for the tilting angle, $\phi$, will be discussed shortly. Therefore, we~write
\begin{equation}
V_{dip}(x)=U\frac{x^2+(n d)^2-3\left[x\cos\phi+n d\sin\phi\right]^2}{\left(x^2+(nd)^2\right)^{5/2}},
\label{eq2}
\end{equation}
where $U$ determines the strength of the dipole-dipole~interaction.  Values of $U$ are in units of $\hbar^2d/m$ unless otherwise noted.

\begin{figure}
\centering
\includegraphics[width=0.7\columnwidth]{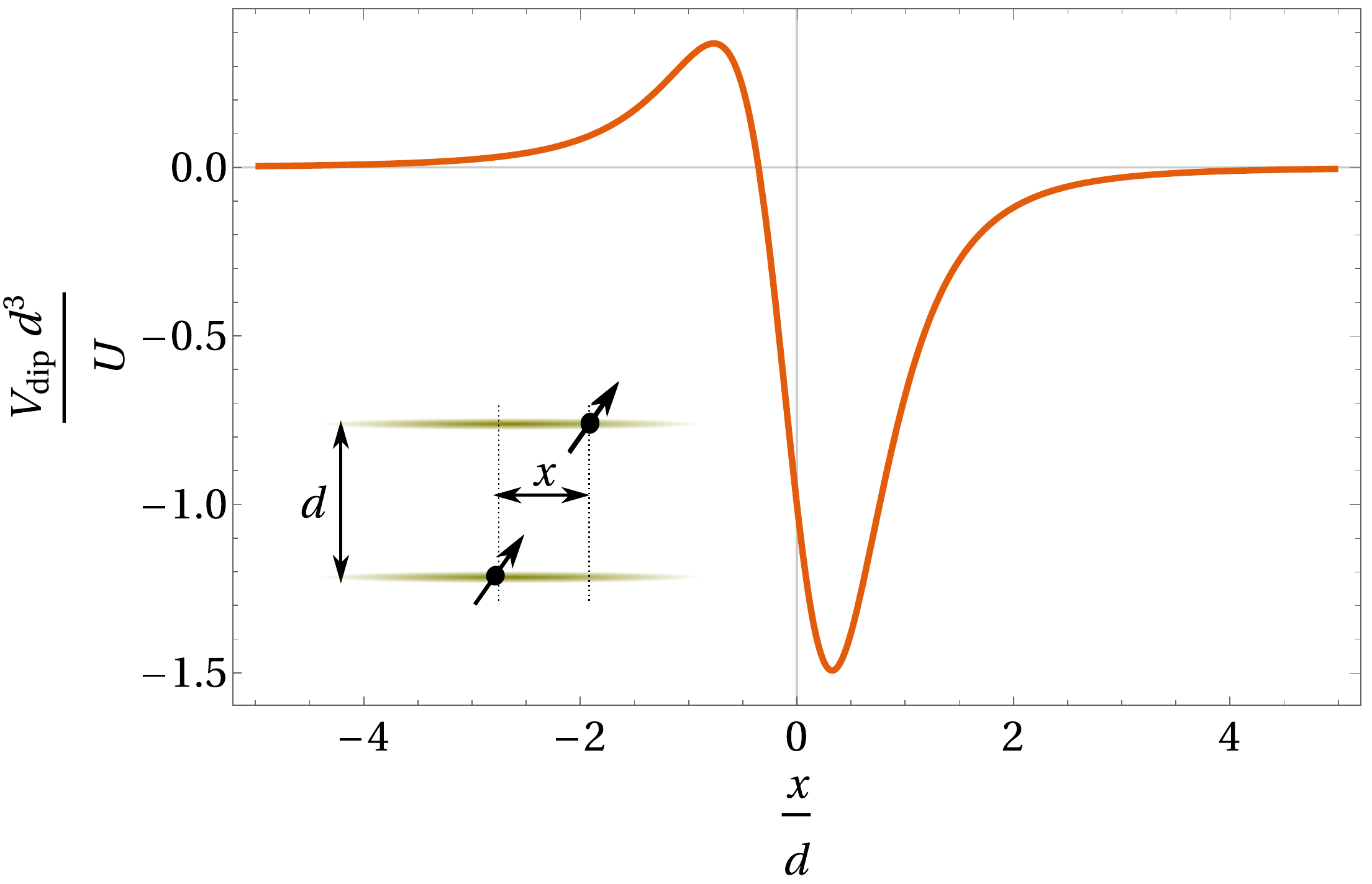}
\caption{{ The interaction potential between adjacent tubes at the ``magic angle" $\phi=\phi_m$ (see the text for the definition of the angle). The~potential is given by Equation~(\ref{eq2}) with $n=1$ and $\phi=\phi_m$: $V_{dip}=-U(d^2+2\sqrt{2}x d)/(d^2+x^2)^{5/2}.$}}
\label{Figure2_new}
\end{figure}

It is complicated, if~not impossible, to~calculate exactly binding energies of $H^{dd}$ for large clusters of dipoles. Instead, we~resort to a harmonic approximation. The~model of choice, coupled quantum harmonic oscillators,
\begin{equation}
H^{osc}=-\frac{\hbar^2}{2m}\sum_i\frac{d^2}{dx_{i,\alpha}^2}+\frac{\mu}{4}\sum_{i,j,\alpha,\beta}\omega_{\alpha \beta}^2(x_{i,\alpha}-x_{j,\alpha}-b_{\alpha\beta})^2+V^{shift},
\label{eq4}
\end{equation}
is exactly solvable~\cite{arms11}.  Here $\mu=m/2$ is the reduced mass of the dipolar particles, $\omega_{\alpha \beta}$ is the coupling frequency between particles in different layers (if $\alpha=\beta$ then $\omega_{\alpha\beta}=0$), $b_{\alpha\beta}$ is the origin shift of the coupling frequency, and~$V^{shift}$ is a constant energy shift. {The parameter $b_{\alpha\beta}$ is present because a~spatially shifted oscillator more accurately reflects $V_{dip}$, as~the minimum of $V_{dip}$, in~general, does not occur at $x=0$ (see Figure~\ref{Figure2_new}).}

The parameters of Equation~(\ref{eq4}) should be adapted to the system of interest depicted in Figure~\ref{Figure1}.  Our philosophy is that the properties of two dipoles in adjacent layers should be reproduced by our oscillator model.  The~dipoles experience an overall attraction for $U>0$ (i.e., $\int V_{dip}(x)\mathrm{d}x<0$), which leads to a two-body bound state in one spatial dimension at any interaction strength~\cite{landaubook,simon1976}. We~would then like to use the energy of this two-body state, as~well as its size, to~determine the parameters of $H^{osc}$: $\omega_{\alpha\beta}$, $b_{\alpha\beta}$ and $V^{shift}$.
These parameters are obtained by variationally solving the exact Hamiltonian from Equation~(\ref{eq3}) for two particles:
\begin{equation}
H^{dd}_2=-\frac{\hbar^2}{2\mu}\frac{d^2}{dx^2}+V_{dip}(x),
\label{eq1}
\end{equation}
where $x$ is the relative in-tube distance between two~dipoles.



To establish the coupling frequency between adjacent layers, $\omega_{12}$, and $b_{12}$ we find the function of the Gaussian form, $\psi\propto \exp\left(-A(x-B)^2\right)$, that minimizes the expectation value of $H^{dd}_2$. We~note that the function $\psi$ is the ground state of the Hamiltonian from Equation~(\ref{eq4}) for two particles, i.e.
\begin{equation}
H^{osc}_{2}=-\frac{\hbar^2}{2\mu}\frac{d^2}{dx^2}+\frac{\mu\omega_{12}^2(x-b_{12})^2}{2}+V^{shift}_2 ,
\end{equation}
whose frequency, $\omega_{12}$, is related to the variational parameter $A$ by $\omega_{12}=2A\hbar/\mu$ and $b_{12}=B$. The~energy shift, $V^{shift}_2$, is used in $H^{osc}_2$ to set the two-body energy at the correct position, $V^{shift}_2=E_2-\hbar\omega_{12}/2$, where $E_2$ is the exact ground state energy of $H^{dd}_2$. { We calculate this energy by solving the Schr{\"o}dinger equation in coordinate space. We~first use a lattice grid to discretize the kinetic energy operator, which leads to a linear system of equations. This system is then solved by matrix diagonalization.} The error can be made arbitrarily small by increasing the number of points used for discretization.  Therefore, all parameters of $H^{osc}$ for two particles are determined.  To~set the interactions that appear in $H^{osc}$ beyond adjacent layers, we~use the scaling properties (see Reference~\cite{arms12}) of the dipole-dipole Hamiltonian~(\ref{eq1}) to adjust the frequencies and shifts:
\vspace{6pt}
\begin{eqnarray}
\omega_{\alpha\beta}=\frac{\omega_{12}(U/\alpha-\beta)}{(\alpha-\beta)^2}\;,\label{w_scale}\\
b_{\alpha\beta}=\frac{b_{12}(U/\alpha-\beta)}{(\alpha-\beta)^2}\;,\label{b_scale}\\
V^{shift}_{2,\alpha\beta}=\frac{E_2(U/\alpha-\beta)}{(\alpha-\beta)^2}-\frac{\hbar\omega_{\alpha\beta}}{2},\label{V_scale}
\end{eqnarray}
where the functions $\omega_{12}(U)$, $b_{12}(U)$, and~$E_2(U)$ describe the dependence on the dipole strength of the frequency, origin shift, and~two-body energy, respectively.
They are obtained by following the variational procedure described above for a set of values of $U$. { The scaling properties are obtained by making the Hamiltonian dimensionless, by~using the inter-layer distance, $d$, as~the unit of length, and~then seeing how the different quantities scale with this distance. The~dimensionless Schr\"odinger equation is
\begin{equation}
-\frac{1}{2}\frac{d^2\psi}{d\bar{x}^2}+\frac{Um}{\hbar^2d}\frac{(\bar{x}^2-2)}{(\bar{x}^2+1)^{5/2}}\psi=\frac{Emd^2}{\hbar^2}\psi,
\end{equation}
where $\bar{x}=x/d$.  From~this equation, it can be seen that the dipole strength scales with $1/d$ and that the energy must be scaled by $1/d^2$, affecting the shift as shown in Equation~(\ref{V_scale}).  Regarding the other two scaling relationships, the~expectation value of a two-body Hamiltonian can be written as
\begin{equation}
E_{exp}=\sqrt{\frac{\pi}{2A}}\int e^{-2A(x-B)^2}\left(-\frac{\hbar^2}{2m}\frac{\partial^2}{\partial x^2}+V_{dip}(x)\right)\mathrm{d}x,
\end{equation}
which means that $E_{exp}=\frac{\hbar^2}{md^2}G(\omega_{12}d^2,b_{12}/d,U/d)$, where $G$ is some function.  Equations~(\ref{w_scale}) and (\ref{b_scale}) follow from the functional form of $E_{exp}$.
}

{ For all our calculations, we~consider the angle, $\phi$, to~be the so-called ``magic angle'', $\phi_m$. This~is the angle where the intra-layer interaction, $V_{dip}$, vanishes, and~is determined by $\cos(\phi_m)=1/\sqrt{3}$ ($\phi_m\approx 54.7^\circ$). The~inter-layer interaction is presented in Figure~\ref{Figure2_new}. To~explain our choice of angle, we~discuss below what happens if $\phi<\phi_m$ or $\phi>\phi_m$.
If $\phi<\phi_m$, then there is attraction between particles within the layers. We~do not consider this case further, because~a many-body system of attractive dipoles collapses, i.e.,~the limit $\lim_{N\to \infty}E_N/N$ is not a finite number (cf.~Reference~\cite{mcguire1964} for bosons interacting with zero-range potentials). We~note that one could stabilize the system of attractive bosons by including a short-range repulsion, see, e.g.,~theoretical works References~\cite{kora2020,palo2020,rzazewski2020} inspired by recent observations of coherent droplets in dipolar systems~\cite{tanzi2019, bottcher2019, chomaz2019,bottcher2019a}. We~do not investigate this possibility here. For~$\phi>\phi_m$, there is repulsion within the same tube and attraction between the tubes.  We modeled this system with two-dimensional layers before~\cite{arms12} with an inverted oscillator representing the repulsion.  While this reasonably modeled how such a system might fall apart, the~inverted oscillator was difficult to constrain, and we sought a more realistic way of representing the repulsion. With~the results found in~\cite{karwowski2008, karwowski2010,arms15,volosniev2019}, showing that we could treat the individual layers separately, we~inserted the exact repulsion in the in-layer energies, coupled with harmonic attraction between the layers.  This treatment failed to agree with earlier SVM calculations in Reference~\cite{volo13}.  For~example,  it was found that even at 55$^\circ$ (just past the ``magic angle''), the~oscillator model had energies that were noticeably different from what was calculated before. This happened because the long-range in-layer dipole-dipole repulsion pushes the particles far from each other into the region, where the harmonic oscillator does not reproduce well the intra-layer attraction.  Therefore, in~the present work, calculations are performed at the ``magic angle'' only. }

Now all parameters of $H^{osc}$ are determined (note that $V^{shift}$ from Equation~(\ref{eq4}) is the sum of all the $V^{shift}_2$ for all pairs). We~can move on to calculating energies of clusters. However, before~that, we~note that there are other ways to determine the parameters of $H^{osc}$. One could, for~example, avoid using $V^{shift}$ and establish frequencies variationally, similar to two-dimensional calculations of Reference~\cite{wang2006}. An~advantage of such an approach would be that the obtained energies rigorously establish an upper bound on the energy. A~disadvantage of neglecting $V^{shift}$ would be that a Gaussian wave function fails to describe weakly bound states. For~example, it predicts a critical value for two-body binding in two spatial dimensions~\cite{wang2006}.
{ One could also estimate the parameters of $H^{osc}$ from $V_{dip}$ (see Figure~\ref{Figure2_new}) using the limit of large $U$, i.e.,~when particles move only in the vicinity of the potential minimum. The~interaction potential close to its minimum can be written as
\begin{equation}
V_{dip}=-U\frac{d^2+2\sqrt{2}x d}{(d^2+x^2)^{5/2}}\simeq  \frac{U}{d^3} \left[- 1.49 + 4.34 \left(\frac{x}{d}-x_{m}\right)^2 - 3.32  \left(\frac{x}{d}-x_{m}\right)^3-5.03 \left(\frac{x}{d}-x_{m}\right)^4\right],
\label{eq:expans}
\end{equation}
where $x_{m}=\frac{3\sqrt{17}-5}{16\sqrt{2}}$ determines the position of the minimum of $V_{dip}$.
This expansion leads~to an~estimate of the ground state energy
\begin{equation}
E_2=-1.49\frac{U}{d^3}+\sqrt{4.34\frac{U}{d^3}\frac{\hbar^2}{md^2}}-0.869\frac{\hbar^2}{md^2},
\end{equation}
the first two terms here are calculated using the first two terms in the expansion in Equation~(\ref{eq:expans}), the~last term is calculated considering the last terms in  Equation~(\ref{eq:expans}) as perturbations. According to this estimate, $V_{dip}$ can be approximated by a harmonic oscillator potential if $-1.49\frac{U}{d^3}+\sqrt{4.34\frac{U}{d^3}\frac{\hbar^2}{md^2}}\gg 0.869\frac{\hbar^2}{md^2}$, which is not satisfied for parameter regimes we consider below. We~leave an exploration of different ways to determine parameters of $H^{osc}$ to future studies. Instead, we~compare the energies from our oscillator model to the exact results.
}

\begin{figure}
\centering
\includegraphics[width=0.8\columnwidth]{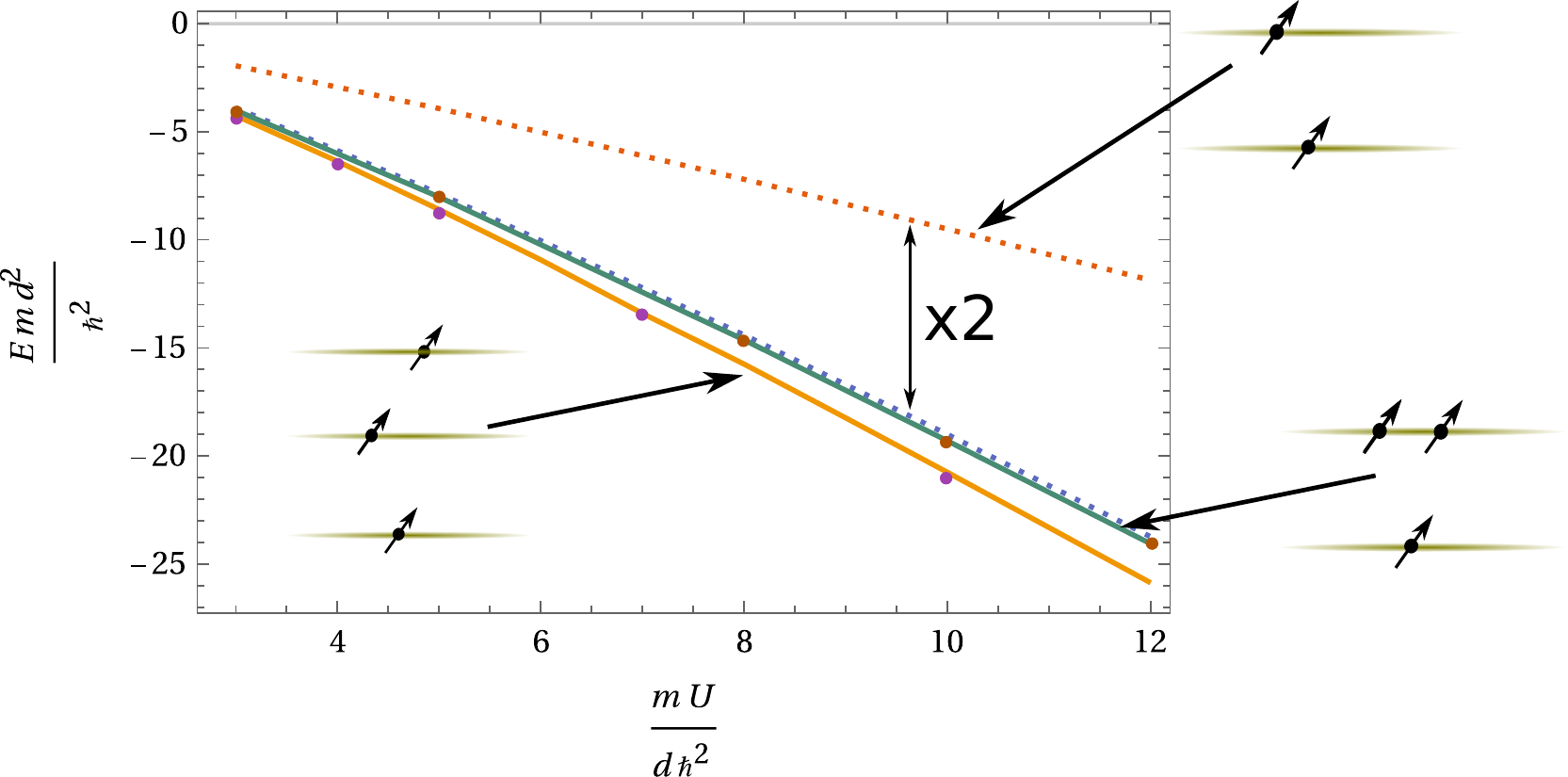}
\caption{Comparison between the harmonic oscillator (HO) model and the stochastic variational method (SVM) in obtaining energies for few-body dipolar clusters. The~solid curves represent the results of the HO model: the upper curve is for the 12 system, the~lower is for the 111 system. The~dots are the corresponding SVM results. The~111 system is slightly more bound than 12 system due to an additional attraction between the outer layers. For~comparison, we~also plot the energy of the 11 system (see the upper dotted curve) whose energy, by~construction, is the same in the HO and SVM calculations. The~lower dotted curve presents two times the energy of the 11~system.}
\label{Figure2}
\end{figure}

For convenience, we~first introduce the labeling for bound states (see Figure~\ref{Figure2}): 11 means a bound state made of two particles in adjacent layers, 12 refers to two particles in one layer and a single particle in the adjacent layer, 111 is a bound state of three particles with one particle per tube, etc.
After determining all parameters of the oscillator model, we~compare the ground state energies of free (no external trap) few-body systems obtained in the oscillator model with results calculated with the stochastic variational method (SVM) (for the description of the method see~\cite{svm_book, mitroy2013}).

These comparisons can be seen in Figure~\ref{Figure2}; we also tabulate energies for certain values of $U$ in Table~\ref{tab1}. The~harmonic oscillator and variationally obtained results are indeed very close in all cases. This is also demonstrated in Figure~\ref{Figure2}, where the results of the two methods are compared.  They agree very well, with~the worse comparisons within about 1.5\%.  Similar comparisons for chains of dipoles can be found in~Reference~\cite{volo13}. For~our further calculations, we~will use only the ground state energies of $H^{osc}$. We~assume that for~a given cluster, the~population of all bound excited states is negligible in comparison to the population of the ground state. { To validate this assumption, we~note that for small values of $U$ there are no excited states. We~find numerically that the first excited bound state for two particles in two adjacent layers appears at $U\simeq 3.7\hbar^2d/m$. The~excited states remain weakly bound for all considered values of $U$. For~example, for~$U=8\hbar^2d/m$, the~ground state is about $7.43$ times more bound than the first excited state. Each weakly two-body state gives rise to a family of shallow few-body states. These states have small binding energies, which allows us to refrain from considering them here. }
Finally, let us give an estimate for the temperature scales that correspond to the calculated binding energies. We~assume that $d=0.5\mu$m and $m=m(^6\mathrm{Li} ^{133}\mathrm{Cs})$, which leads to $E_2\simeq 50 nK\times k_B$ for $U=5$, where $k_B$ is the Boltzmann constant. Smaller values of $U$ require even smaller temperatures for observation of few-body clusters, therefore, in~what follows we focus on $U=5$ and $U=8$.


\unskip


\begin{table}
\caption{Comparison between the harmonic oscillator (HO) model and the stochastic variational method (SVM) in obtaining energies for few-body dipolar clusters.  The~different clusters are three and four particle chains (labeled 111 and 1111, respectively), and~a system with two particles in one layer and a single particle in the adjacent layer (labeled 12).  The~units of energy are $\hbar^2/(md^2)$, and~the units of $U$ are $\hbar^2d/m$. }
\begin{center}\begin{tabular}{  c  c  c  c c  c c }
\boldmath{$U$} & \textbf{111 (HO)}
& \textbf{111 (SVM)} & \textbf{1111 (HO)} & \textbf{1111 (SVM)} &\textbf{12 (HO)} & \textbf{12 (SVM)} \\
3 & $-$4.35 & $-$4.45 & $-$6.88 & $-$7.11 & $-$4.02 & $-$4.01\\
5 & $-$8.67 & $-$8.81 & $-$13.63 & $-$13.95 & $-$8.04 & $-$8.02 \\
10 & $-$20.67 & $-$20.97 & $-$32.33 & $-$32.96& $-$19.31   & $-$19.30  \\
\end{tabular}
\end{center}

\label{tab1}
\end{table}

\section{Abundances of~Clusters}
\label{subsec:part}

We consider five layers, each with two dipolar molecules (particles) inside.
We assume that every particle is in the harmonic oscillator, $m \omega_0^2 x_{i,\alpha}^2/2$. This can be either due to an external trapping potential, or~a way to simulate a finite density of a many-body system. We~then calculate the fractional occupancy of given clusters as a function of temperature.  For~simplicity, we~assume that the particles are distinguishable, thus they obey Boltzmann statistics. { We discuss this assumption in detail at the end of the next section.}

Few-body clusters range from the simplest, a~two-particle bound state of particles in adjacent layers, up~to a bound state of all ten particles.  We also include the possibility that all ten remain unbound, in~which case the energies are approximately given by the energies of the states of the confining harmonic trap of the layer.  The~fractional occupancy of any state $k$ is
\begin{equation}
f_k=\frac{e^{-\beta E_k}}{Z}, \label{fraction}
\end{equation}
where $E_k$ is the energy of state $k$, $Z$ is the partition function and $\beta=1/k_BT$.  The~partition function in the canonical ensemble is
\begin{equation}
Z=\sum_kg_ke^{-\beta E_k},
\end{equation}
where $g_k$ is the degeneracy of the $k$th energy. We~write the energy of the various cluster configurations as a sum of two components:
\begin{eqnarray}
E_k&=& \sum_{\mathrm{bound}} \left[\epsilon_j+\hbar\omega_0 \left(n^{CM}_j+\frac{1}{2}\right)\right]+\hbar\omega_0\sum_{\mathrm{free}}\left(n_{l}+\frac{1}{2}\right).\nonumber\\
&=&\mathcal{E}_k+\left(\mathcal{N}_\nu+\frac{\nu}{2}\right)\hbar\omega_0.\label{energy}
\end{eqnarray}

\begin{figure}
\centering
\includegraphics[width=0.5\columnwidth]{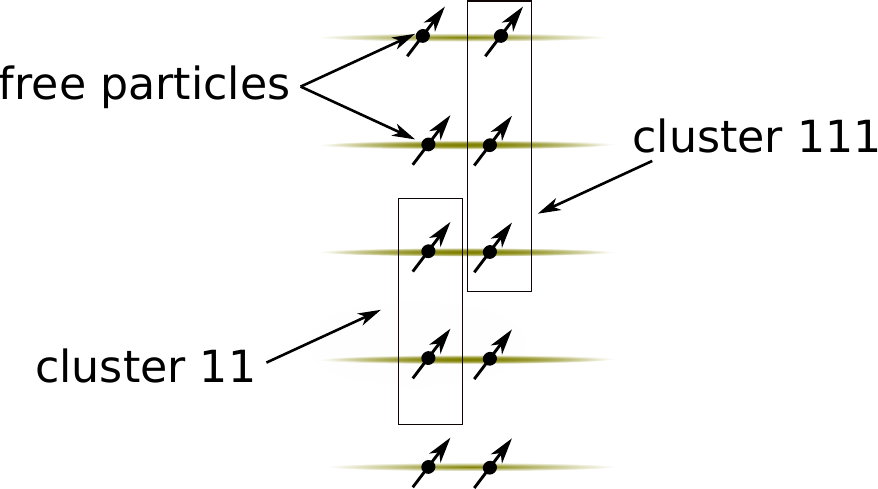}
\caption{A figure to illustrate Equation~(\ref{energy}). This specific configuration has five free particles, and~two clusters. The~energy of each cluster consists of two parts: The binding energy, which is calculated as in Section~\ref{subsec:energ}, and~the center-of-mass part, which is determined by the confining harmonic~oscillator.}
\label{Figure2a}
\end{figure}

The first component, $\mathcal{E}_k=\sum_{\mathrm{bound}} \epsilon_j$, is the binding energy of all clusters in the state $k$, with~$\epsilon_j$ being the binding energy of the $j$th cluster in the configuration $k$.  The~first line in Equation~(\ref{energy}) also contains sums over all the various oscillator degrees of freedom in the configuration.  The~free particles are particles moving in the oscillator potential defined by $\omega_0$, with~the corresponding energy levels and quantum numbers, $n_l$.  The~center of masses (CM) of the cluster(s) also have the same spectrum. To~simplify notation, we~introduce the quantum number $\mathcal{N}_\nu \equiv \sum_{\mathrm{bound}} n^{CM}_j+ \sum_{\mathrm{free}}n_{l}$. The~value of $\nu$ defines how many oscillator degrees of freedom we have. To~illustrate the decomposition of the energy $E_k$, let us consider the configuration $k$ presented in Figure~\ref{Figure2a}. This configuration has two clusters and five free particles. The~energy $\mathcal{E}_k$ is the sum $\epsilon_{11}+\epsilon_{111}$. The~value of $\mathcal{N}_\nu$ can take any integer value. It is given by the decomposition $\mathcal{N}_{\nu=7}=\sum_{i=1}^7 n_i$, where $n_i=0,1,2...$.


The energy, $\epsilon_j$, is obtained by using the harmonic approximation from the previous section. By~construction, $\epsilon_j$ is not affected by the harmonic oscillator potential, $m \omega_0^2 x_{i,\alpha}^2/2$, whose length is much larger than the size of the cluster.
Please note that to write the energy $E_k$, we~assume that the cluster-cluster and cluster-(free particle) interactions are negligible. This assumption relies on the two observations: (i) by construction, the~harmonic oscillator length is much larger than the range of the dipole-dipole potential, therefore, for~low-lying excited states we may approximate $V_{dip}$ with a~zero-range interaction; (ii) the interaction due to a short-range potential can shift the energy by only about  $\hbar \omega_0$. This statement relies on comparing the energies in a~weakly interacting limit to that of a~strongly interacting limit for zero-range interaction models, see, e.g.,~References~\cite{busch1998, dehkharghani2016}. This shift is not important for our qualitative~discussion.

For a specific state, the~CM and the free motion would also appear with specific quantum numbers.  Since we are primarily interested in which specific clusters are prevalent at a given temperature, we~then include all possible oscillator excitations by summing them up, so that the probability of a~specific cluster configuration, $F_k$, is given by summing $f_k$ from Equation~(\ref{fraction})
\begin{equation}
F_k=\frac{g_k}{Z}e^{-\beta\left(\mathcal{E}_k+\frac{\nu_k\hbar\omega_0}{2}\right)}\left(\frac{1}{1-e^{-\beta\hbar\omega_0}}\right)^{\nu_k},
\end{equation}
where $\mathcal{E}_k$, $\nu_k$, along with the degeneracy $g_k$ must be
determined for each cluster configuration.  The~partition function, extracted from the condition that $\sum f_k=1$, is then
\begin{equation}
Z=\sum_kg_ke^{-\beta\left(\mathcal{E}_k+\frac{\nu_k\hbar\omega_0}{2}\right)}\left(\frac{1}{1-e^{-\beta\hbar\omega_0}}\right)^{\nu_k}.
\end{equation}

As an example, consider the cluster configuration in Figure~\ref{Figure2a}. We~have
$\nu=7$, since there are five free particles, the~CM of the 11 cluster, and~the CM of the 111 cluster. The~binding energy of the clusters is given by $\mathcal{E}_k=\epsilon_{11}+\epsilon_{111}$. There are 176 ways to distribute the clusters 11 and 111 among five different layers. Therefore $g_k=176$, and~furthermore, the~probability to find a configuration with a single 11 cluster and a single 111 cluster is given by
\begin{equation}
F_{11+111}=\frac{176}{Z}e^{-\beta(\epsilon_{11}+\epsilon_{111}+\frac{7}{2}\hbar\omega_0)}\left(\frac{1}{1-e^{-\beta\hbar\omega_0}}\right)^7,
\end{equation}
where we introduced the convention that $A+B$ means that the clusters $A(=11)$ and $B(=111)$ exist simultaneously in the~system.

\section{Results and~Discussion}
\label{sec:Res}

Numerical applications of the formulation quickly contain many
configurations.  In~the present report, we~restrict ourselves to two
relatively simple systems, yet sufficiently complicated to reveal
general~features.

\subsection{One Particle per~Tube}

The first system considered is five layers each with one particle.
There are seven different clusters in this system: chains of five,
four, three, and~two particles, five free particles, two separate
chains of two particles each, and~finally a chain of three particles
separate from a chain of two particles.  We do not consider clusters
consisting of four particles when the middle layer is empty, because~our
clusters should at least have one particle in a layer linking them
together. Otherwise they are very weakly bound and effectively
separate~structures.

\begin{figure}
\centering
\includegraphics[width=0.95\columnwidth]{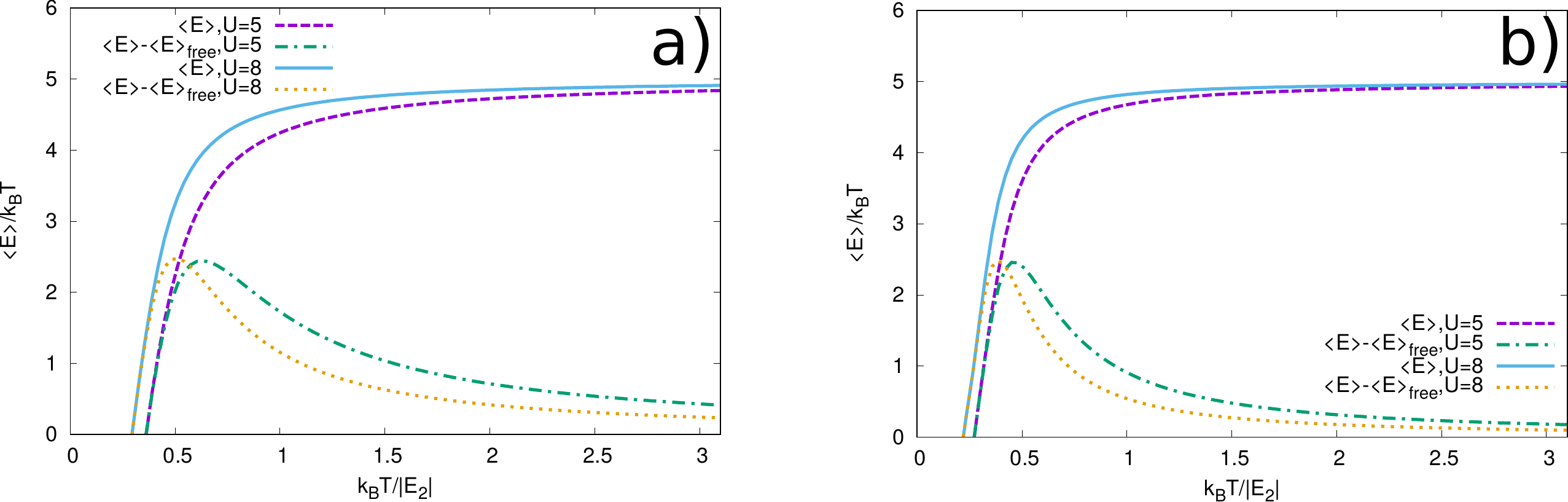}
\caption{This plot shows the energy of the one-particle-per-layer system, $\langle E\rangle$, and~the difference of the energy with the energy of the system of completely free particles, $\langle E\rangle-\langle E\rangle_{free}$.  Panel ({\textbf{a}}) shows curves for different $U$ with oscillator length values being twice the inter-layer distance, and~in panel ({\textbf{b}}) the oscillator length is $\sqrt{10}$ the inter-layer distance.  { The energy curves all start at large negative energies because of the finite binding energy at low temperatures, then approach the high-temperature limit of 5 $k_BT$ (the equipartition of energy limit for this system)}.  The~curves showing the difference of energies start to separate from the energy curve at $k_BT/E_2\approx$0.25 to 0.4 as the free-particle state becomes populated before rapidly turning over and descending towards the high-temperature limit of~0.}
\label{Figure 8}
\end{figure}

The energy of the system as function of temperature is first
calculated as the Boltzmann weighted average over cluster
configurations. The~results are shown in Figure~\ref{Figure 8} for
different interactions and trap frequencies.  The~overall behavior of
the energies is not surprisingly a move from the ground state values
at low temperature to the high-temperature limit of $5k_B T$ for five
free particles in the present system.  This limit is seen by comparing
to the temperature dependent average energy of five free particles.
This limit is almost reached at a temperature of about the dimer
bound-state energy of two particles in adjacent~layers.

\begin{figure}
\centering
\includegraphics[width=0.95\columnwidth]{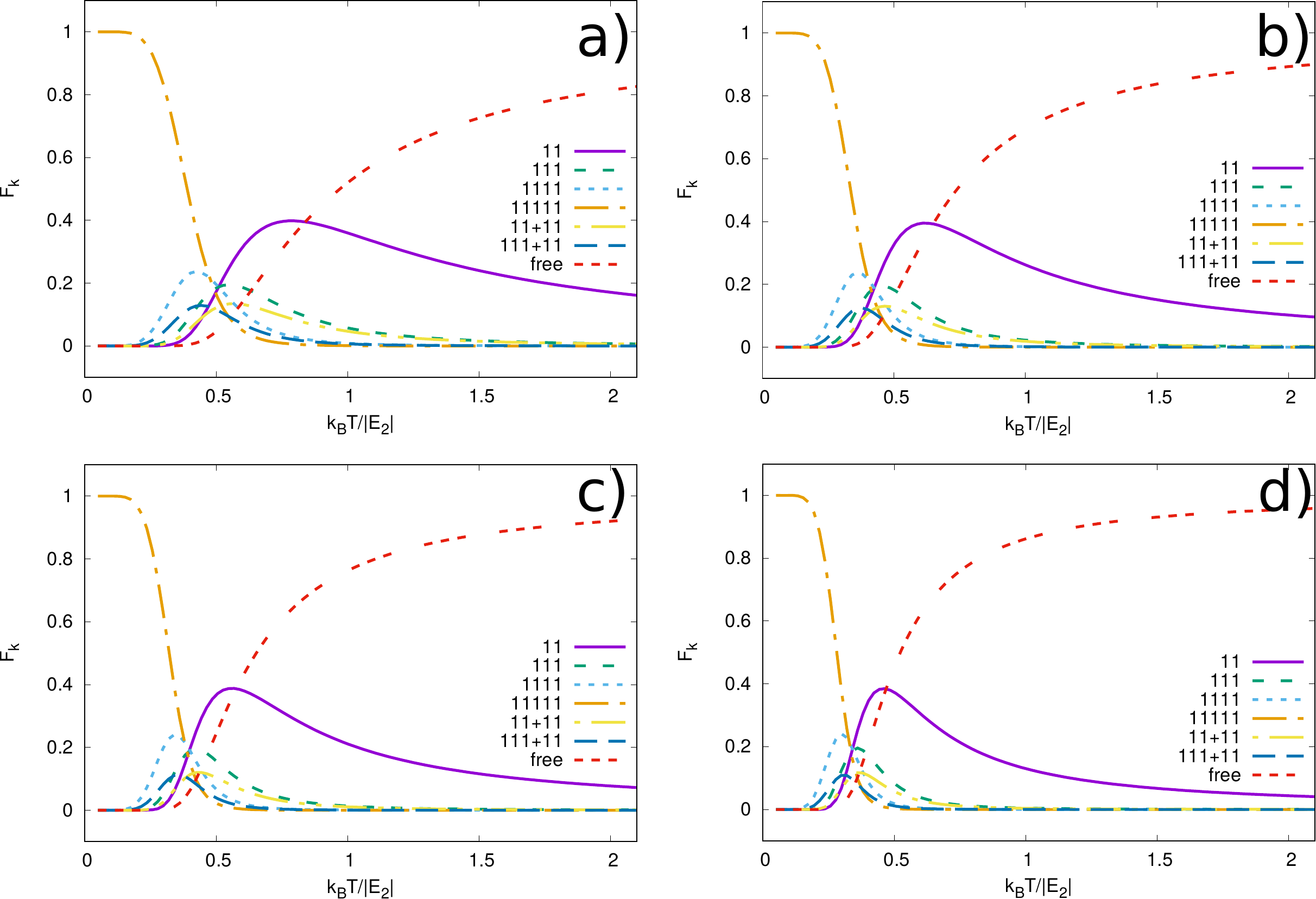}
\caption{Fractional occupancies of different clusters as a function of temperature for a system of five 1D layers each with one particle. (\textbf{a}): The confining frequency is chosen such that the oscillator length in the tubes is equal to twice the inter-layer distance. The~dipole strength is $U=5$. (\textbf{b}): The confining frequency is chosen such that the oscillator length in the tubes is equal to twice the inter-layer distance. The~dipole strength is $U=8$. (\textbf{c}) The dipole strength is $U=5$, and~the confining frequency is chosen such that the oscillator length in the tubes is equal to $\sqrt{10}$ the inter-layer distance. (\textbf{d}): The dipole strength is $U=8$ and the confining frequency is that same as in panel (\textbf{c}).}
\label{fig1}
\end{figure}

Results for fractional occupancies are shown in Figure~\ref{fig1}a, where we see that the dominant cluster is the fully bound
5-particle chain at low temperature.  Its occupancy decreases rather
quickly from unity to zero, and~as the temperature increases, the~less-bound structures appear.  The~free-particle occupancy increases
steadily as expected towards unity at high temperature.  The~free particles are
already dominating at intermediate temperature, where the second
largest contribution consists of bound dimer~systems.

In Figure~\ref{fig1}b, the~interaction strength increased to
$U=8$ in comparison to Figure~\ref{fig1}a, but~the plots are very
similar, since the $x$-axis is scaled by the two-body binding energy.
In both cases, at~just under $k_BT/E_2=0.5$, we~see the most mixed
system with most of the clusters having significant abundance.  None
have a fraction greater than about 0.2 at this temperature. Please note that the $U=5$ plot appears to show
slightly longer tails into higher~temperatures.

Figure~\ref{fig1}c changes the confinement frequency.  This effectively changes the density of the particles in a tube, since the oscillator length of the tube, $b=\sqrt{\hbar/m\omega_0}$ is changed by changing $\omega_0$.  For~the sake of argument, we~relate this length to the distance between the layers, obtaining the relationship $\omega_0\propto1/(\alpha^2d^2)$, where $\alpha$ is a scaling factor that can be experimentally controlled.  In~Figure~\ref{fig1}c we then take $\alpha=\sqrt{10}$, which decreases the density of the layer, while keeping the interaction strength the same as in Figure~\ref{fig1}a. The~primary effect is similar to increasing the interaction strength that is shifting the emergence of smaller fragments to smaller~temperatures.

This is emphasized in Figure~\ref{fig1}d, which has both $U=8$ and decreased density, and~the ``melting'' of the fully bound cluster occurs at the smallest temperature.  The~relative maxima of the curves remain of a similar height in all the plots.  Therefore, interaction strength and confinement frequency can cause similar movements on the temperature scale. { It may appear from glancing at Figure~\ref{fig1} that the interaction strength does not have a large effect on our results.  Recall, however, that the $x$-axis has been scaled by the two-body energy, which is greatly influenced by the interaction strength.  The~figures would be quite different without this scaling.  The~re-scaling makes it clear that the two-body energy sets the relevant energy scale for the system.}

\subsection{Two Particles per~Tube}

When we include two particles per tube, the~number of clusters
increases dramatically.  There are 119 non-degenerate cluster
configurations, and~we do not consider any clusters where there is an~empty tube between different members of the cluster.  An~intermediate
attractive ingredient is again needed to provide an effectively bound
system in contrast to separate~configurations.

\begin{figure}
\centering
\includegraphics[width=0.8\columnwidth]{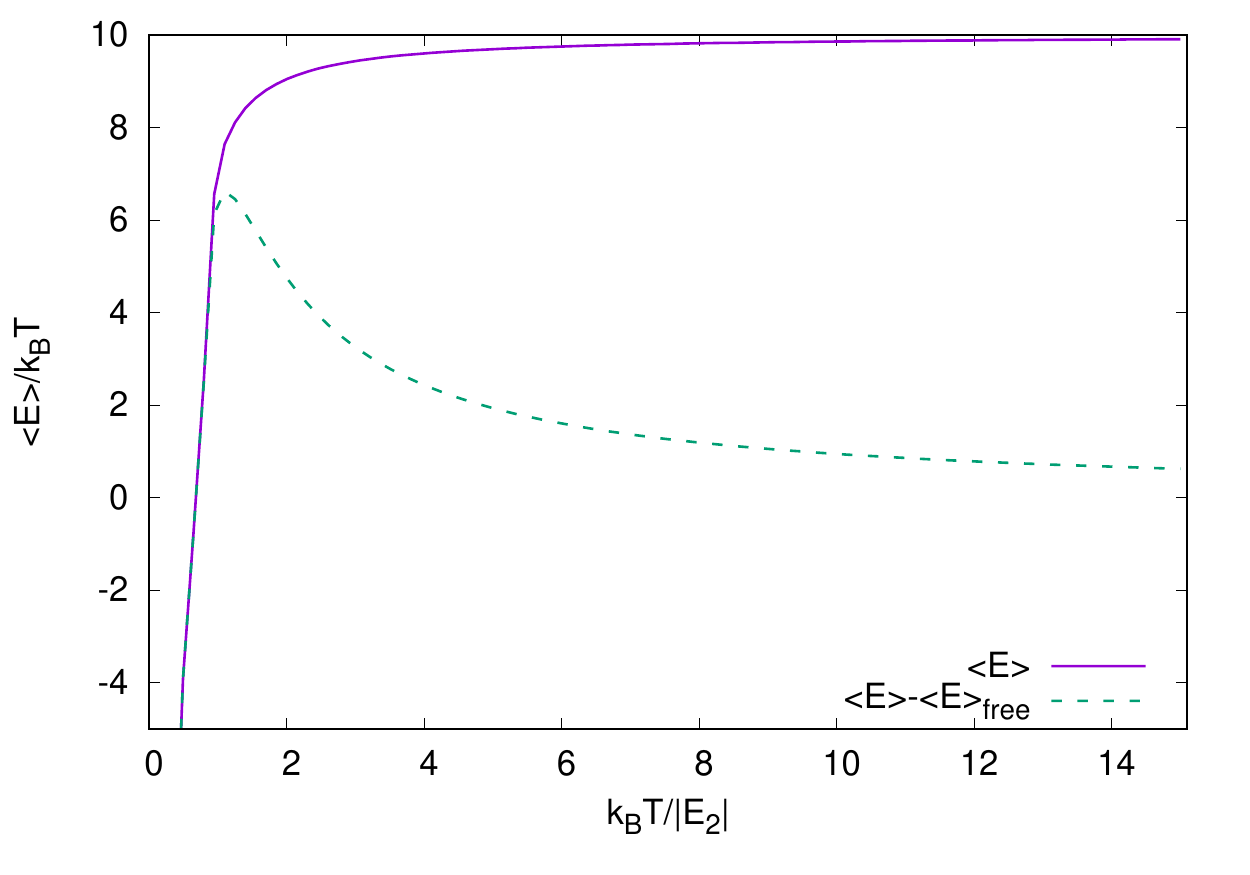}
\caption{This plot shows the same as in Figure~\ref{Figure 8}, but~for two-particle-per-layer system.  { The energy curve again starts at large negative energies, then approaches the high-temperature limit of 10 $k_BT$ (the equipartition limit for 10 total particles in 1D harmonic potentials)}.  The~curve showing the difference of energies starts to separate from the energy curve at $k_BT/E_2\approx$1 as the free-particle state becomes populated before rapidly turning over and descending towards the high-temperature limit of~0.  In~contrast with the single particle per layer system, the~high-temperature limits are achieved much more slowly.  The~interaction strength for this plot was $U=5$.}
\label{Figure 9}
\end{figure}

In Figure~\ref{Figure 9} we show the average energy compared to the
energy of ten free particles as function of temperature for one
interaction and one trap frequency.  This energy again increases from
the bound-state value to the high-temperature result, $10k_B T$, for~ten free particles as we have in this system.  The~transition is
almost achieved at a temperature of about twice the dimer binding of
two particles in adjacent~tubes.

\begin{figure}
\centering
\includegraphics[width=0.95\columnwidth]{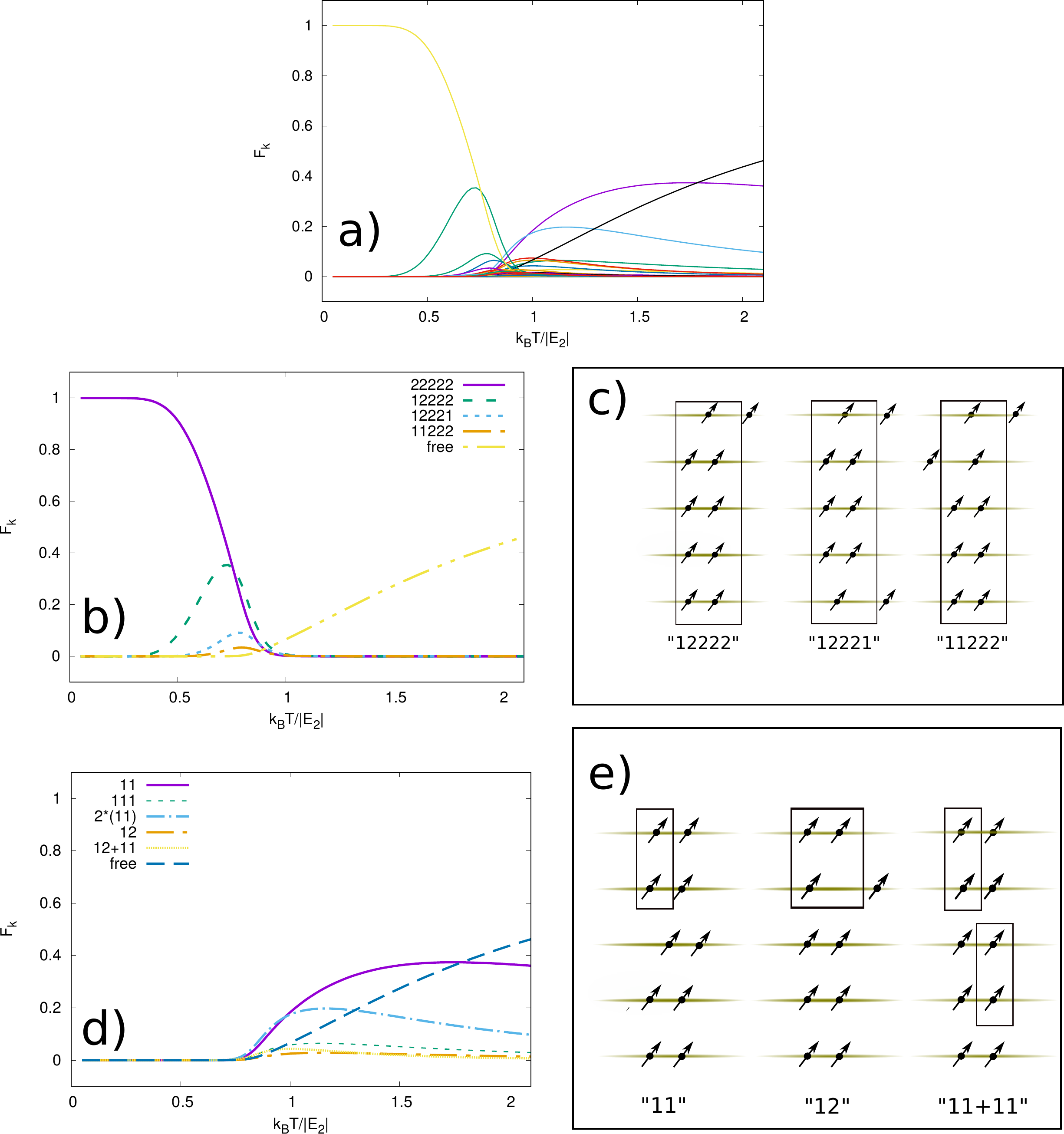}
\caption{{(\textbf{a})}  Fractional occupancies of all the clusters as a function of temperature for a system of five 1D layers each with two particles.  The~dipole strength is $U=5$, and~the confining frequency is chosen such that the oscillator length in the tubes is equal to twice inter-layer distance.  The~clusters are labeled by their layer occupancy, so a cluster consisting of one particle each in adjacent layers would be labeled '11'.  (\textbf{b}) Fractional occupancies of the most populated of the most bound clusters as a function of temperature for a system of five 1D layers each with two particles.  The~dipole strength is $U=5$, and~the confining frequency is chosen such that the oscillator length in the tubes is equal to twice the inter-layer distance. Pictures of the clusters can be seen in (\textbf{c}). (\textbf{d})
Fractional occupancies of the least bound clusters as a function of temperature for a system of five 1D layers each with two particles.  The~dipole strength is $U=5$, and~the confining frequency is chosen such that the oscillator length in the tubes is equal to twice the inter-layer distance.   (\textbf{e}) Pictures of the different smaller clusters with the lowest binding energies.}
\label{fig6}
\end{figure}

Figure~\ref{fig6}{a} shows the occupancies of all the clusters.
There are not too many clusters where we have significant occupation.
Perhaps eight of the 119 clusters can be distinguished in the
figure, where most are too small to be seen.  But~again the ground
state decreases rapidly from unity to zero whereas the free-particle
configurations grow up and dominate at high temperature.  In~comparison with the above system of one particle per tube, the~most
bound clusters dominate to higher temperatures than before, with~the
most mixed system occurring around $k_BT/E_2$ = 0.9.  With~two particles per tube, the~energy gap between the completely bound
cluster and the next clusters is large which means a~higher
temperature is necessary to create other~clusters.

We can take a closer look in Figure~\ref{fig6}{b} which shows the
most bound clusters at low temperatures.  There the sparse amount of clusters
is clear and only the two most bound configurations have
large occupancies before the smaller clusters start to dominate (these
clusters are pictured in Figures~\ref{fig6}c).
Figure~\ref{fig6}d shows the most bound clusters at larger temperatures, and~their
fractional occupancies as a~function of temperature (and pictured in
Figure~\ref{fig6}e).  All the small clusters or collections
of clusters (including the completely free system) start to grow in
occupancy around $k_BT/E_2$ = 0.75.  Only the very least bound, the~free, 11, 12, and~2*(11) clusters achieve significant fractions, while
the rest points back to zero fractional occupancy.  In~general, since
there are so many more clusters or collection of clusters, there are
few clusters that have occupancies >10\%.

\begin{figure}
\centering
\includegraphics[width=0.95\columnwidth]{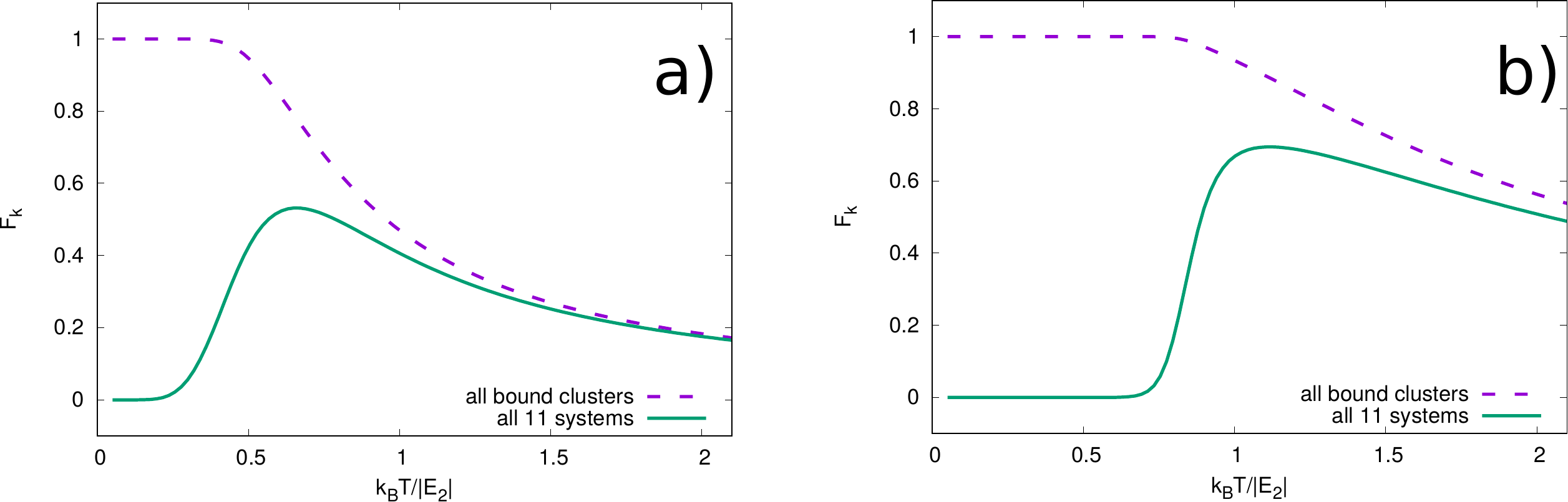}
\caption{  This figure shows the fraction of states with at least one bound cluster of any kind ('all bound clusters') or the sum of all clusters that contain at least one 11-cluster (all 11 systems).  The~dipole strength is $U=5$, and~the confining frequency is chosen such that the oscillator length in the tubes is equal to twice the inter-layer distance.  Panel (\textbf{a}) shows the one-particle-per-layer system and panel (\textbf{b}) shows the two-particle-per-layer~system.}
\label{Figure7}
\end{figure}

Our final Figure~\ref{Figure7} shows two curves in each panel, both
are the sum of cluster occupancies for one and two particles per tube
in panel ({\bf a}) and ({\bf b}), respectively.  The~lower curve in each panel
shows the fraction of all clusters which
contain at least one dimer, while the upper curve shows a related
quantity: the fraction of systems with at least one bound cluster.
In panel (\textbf{a}), the~lower curve is flat until $k_BT/E_2$~=~0.2, then rises rapidly before turning over and declining at the higher temperatures.  The~upper curve is unity until $k_BT/E_2$ = 0.4, then declines, and~with the higher temperatures it approaches the lower curve.  Thus, nearly all the bound systems contain a bound dimer~systems.

In panel ({\bf b}), as~we saw in the previous results, there is little
change in this two-particle-per-tube system, until~higher temperatures
than in the previous single particle case.  The~lower curve, shows
nothing until about $k_BT/E_2$ = 0.7, then rises dramatically before
turning over and declining gradually.  The~upper curve does not begin
to decline rapidly around $k_BT/E_2$ = 0.8.  Again, the curves approach
each other, showing that all bound systems contain a bound dimer at high
temperatures, which is even more clear in the single particle per layer~graph.

{ It is worthwhile noting that the results we discuss here will not change appreciably for bosonic particles, even though we have chosen to work with Boltzmann statistics. Please note that quantum statistics play a role only for particles in the same tube, therefore, when we say bosons we imply particles in the same tube. The~Bose statistics for particles is important {(i)} either when two particles in a tube are a part of a cluster configuration, e.g.,~of a 12 few-body cluster; {(ii)} or when the temperature is below the temperature for condensation. If~there are no symmetry requirements, then the energy of a few-body cluster is minimal when the wave function is symmetric with respect to exchange of two particles in the same tube. Therefore, the~few-body clusters have bosonic symmetry, and~we should not discuss the item {(i)}. The~condensation temperature for our system can be estimated to be $\sim \frac{\hbar^2}{2 m b^2}\times k_B$.
This~temperature is much smaller than the temperature for the melting of a many-body state $\sim \frac{\hbar^2}{2 m d^2}\times k_B$, because~the system has $b^2\gg d^2$ by construction. We~can treat the temperature for condensation as being zero in our work, which allows us to not consider the item {(ii)} further.  This line of argument shows that our results can be used to describe systems of bosons.}

\section{Summary and~Conclusions}
\label{sec:concl}

We study theoretically the temperature dependence of structures of
dipoles trapped in equidistantly separated tubes.  The~dipoles are
tilted by an external field to the ``magic angle,'' where the in-tube
interaction is zero.  The~input that determines the probability to observe
a few-body cluster at a given temperature is the set energies of the
many different cluster configurations.  To~calculate these energies,
we design an accurate method based on an oscillator approximation.  We
demonstrate the validity of the method, and~apply it to calculate the
energies of the different cluster configurations, and~in turn to
obtain the partition function as function of~temperature.

We choose two rather simple systems to be studied in detail as
function of temperature.  The~two systems have five tubes each with
either one or two dipoles.  We first calculate the temperature
dependence of the average energies for different interactions and trap
frequencies in comparison with the energies of the free-particle
system. These dependencies are all qualitatively the same, i.e.,
changing from bound-state values to high-temperature statistical
equilibrium values.  However, finer details reveal weak dependence
on the strength of interaction and the trapping~frequency.

The number of different cluster configurations is relatively large
even for the simple systems we choose to study here.  The~more detailed
results of individual cluster occupancies are available through the
partition function.  We obtain the occupancies of the clusters by
increasing the temperature from zero to much higher than the energy of
a dimer formed by two dipoles in adjacent~layers.  These~occupancies
show a change of the system from the corresponding ground state
towards entirely free particles. However, the~details of this melting
at moderate temperatures reveal how this process proceeds through
intermediate configurations of various clusters.  At~temperatures
around the dimer energy the configurations in the ensemble are mixed
more than at any other temperature. Our~findings show that even though
there are many few-body clusters, most of them are unlikely to be detected in
a many-body system. Indeed, the~system shows a fast transition from a many-body
state at low temperatures to a high-temperature state where only the simplest
clusters (e.g., a~dimer) play a~role. This observation suggests that
effective theories that include only free particles and dimers can accurately describe
the system down to $T\simeq |E_2|/k_B$.

In conclusion, we~have presented a method and derived results for the
melting of one-dimensional systems of relatively few dipoles.
The cluster structures
are clearly very important in systems of many particles at moderate
temperatures.
This suggests a tool for investigating the transition from few- to many
physics by changing the temperature in cold-atom~systems.

In the future, it will be interesting to extend our results to more complicated systems that
could have more particles and/or more tubes. For~a more realistic calculation one should include a short-range intra-layer repulsive interaction even for dipoles at a ``magic angle''. This interaction will decrease the probability to observe few-body structures that have more than one dipole per layer. Please note that an inclusion of a short-range interaction in Equation~(\ref{eq4}) with at most two particles per layer still leads to a solvable model~\cite{arms15}.
One could study as well two-dimensional systems of layers of particles, which are known to support various few-body bound states~\cite{santos2010, volosniev2011, petrov2019}. To~increase the probability to observe non-chain few-body clusters, one should again consider tilting dipoles.
It is impossible to find an angle that turns off completely the dipole-dipole interaction in a two-dimensional layer, which significantly complicates the~problem.

\vspace{6pt}


  \section{Acknowledgments}
  The authors acknowledge the Independent Reseach Fund Denmark (DFF) which supported this work, and the hospitality of Aarhus University (J.R.A. and A.G.V.) where part of this research was done.
  This work also received funding from the DFG Project No. 413495248 [VO 2437/1-1] and European Union's Horizon 2020 research and innovation programme under the Marie Sk\l{}odowska-Curie Grant Agreement No. 754411 (A.G.V.).



\end{document}